\documentclass[nocompress]{spie}  

\usepackage{amsmath,amsfonts,amssymb}
\usepackage{graphicx}
\usepackage[colorlinks=true, allcolors=blue]{hyperref}
\usepackage{lscape}
\usepackage{rotating}
\usepackage{enumerate}
\usepackage{longtable}

\usepackage{soul}
\usepackage[dvipsnames]{xcolor}

\title{The Gaia grid of spectro-photometric standard stars}

\author[a]{N. Sanna}
\author[a,b]{E. Pancino}
\author[b,c]{G. Altavilla}
\author[b,c]{S. Marinoni}
\author[a]{M. Rainer}
\affil[a]{INAF -- Osservatorio Astrofisico di Arcetri, Largo E. Fermi 5, 50125, Florence, Italy}
\affil[b]{Space Science Data Center -- ASI, Via del Politecnico SNC, 00133 Roma, Italy}
\affil[c]{INAF -- Osservatorio Astronomico di Roma, Via Frascati 33, 00078, Monte Porzio Catone (Roma), Italy}

\authorinfo{Author Email: nicoletta.sanna@inaf.it}

\begin{document} 
\maketitle

\begin{abstract}
We describe the preliminary results of a ground-based observing campaign aimed at building a grid of approximately 200 spectro-photometric standard stars (SPSS), with an internal $\simeq 1$\% accuracy (and sub-percent precision), tied to CALSPEC Vega and Sirius systems within $\simeq 1$\%, for the absolute flux calibration of data gathered by {\it Gaia}, the European Space Agency (ESA) astrometric mission. The criteria for the selection and a list of candidates are presented, together with a description of the survey's strategy and the adopted data analysis methods. All candidates were also monitored for constancy (within $\pm 5$ mmag, approximately). The present version of the grid contains about half of the final sample, it has already reached the target accuracy but the precision will substantially improve with future releases. It will be used to calibrate the {\it Gaia} (E)DR3 release of spectra and photometry. 
\end{abstract}


\section{INTRODUCTION}
\label{sec:intro}  
{\it Gaia} is the ESA astrometric, photometric and spectroscopic public survey mission that is repeatedly scanning the entire sky to measure parallaxes, proper motions, and magnitudes for more than $10^9$ stars down to the G $\simeq$ 20.7~mag, along with radial velocities for a bright sub-sample ($\sim$7 million stars with $4\leq G\leq13$~mag).
Five instruments work on board: the Sky Mapper, the Astrometric Field, the Blue Photometer, the Red Photometer and the Radial Velocity Spectrometer [\citenum{2016gaiacollaboration}]. The calibration of the data is quite complex and basically split in two parts: an internal relative calibration (based on {\it Gaia} data only) aimed at reporting the data to the same instrumental reference flux [\citenum{carrasco16}], and an external absolute calibration (based on a relatively small number of external flux calibrators) which reports the instrumental flux in physical units ([\citenum{pancino12}]; Riello et al. in press, Pancino et al. in preparation). In order to externally calibrate the {\it Gaia} data, a grid of spectro-photometric standard stars (SPSS) is needed. 
Here we describe the current preliminary release (V2) of the SPSS grid used to calibrate the spectra and photometry contained in the early and the full data release 3 (EDR3; DR3), foreseen for December 2020 and first half of 2022, respectively\footnote{\url{https://www.cosmos.esa.int/web/gaia}}.
The paper is organized as follows: the SPSS sample, the observations and the spectra reductions are described in Section~\ref{sec:obs}; Section~\ref{sec:v2} presents the current V2 release; the validation of the grid and the conclusions are discussed in Section~\ref{sec:val} and Section~\ref{sec:conc}, respectively.

\section{SPSS candidates and observations}
\label{sec:obs}
A detailed description of the selection criteria adopted for the SPSS grid is reported in [\citenum{pancino12}]. Briefly, the initial requirements on the external calibration of {\it Gaia} are $\simeq$1\% in precision and $\simeq$1--3\% in accuracy.
Several homogeneously calibrated stars ($\sim 200$) of different spectral types (from O to M, including many white dwarfs) covering the  entire {\it Gaia} wavelength range (330--1050~nm) are needed.
None of the already existing standard catalogs met all these requirements and this is why a new dedicated campaign of ground-based observations for the {\it Gaia} SPSS grid started in 2006. For this purpose, photometric and spectroscopic data have been acquired using 8 different telescopes, 4 of them with spectroscopic capabilities too (see [\citenum{altavilla15}];[\citenum{marinoni16}], Pancino et al. in preparation). The targets are spread in magnitude (9$\lesssim$V$\lesssim$15~mag) and sky distribution to be observable from both hemispheres with 2--4\,m class telescopes. The spectra's requirements are resolution R=$\lambda/\delta\lambda\simeq$1000, and S/N ratio $>$100 over most of the spectral range.
Moreover, a short-term flux variations smaller than $\pm$0.5\% (i. e. amplitude $\leq\pm$0.005~mag) is required for all the selected SPSS [\citenum{marinoni16}].   

\begin{table}
 \begin{center}
  \begin{tabular}{|c|c|c|c|c|c|c|}
  \hline
Instrument          & Site               & blue grism       & red grism  & third grism  \\
 &&(range[nm])&(range[nm])&(range[nm])\\
 \hline             
EFOSC2@NTT\,3.6m    & La Silla, Chile   &         \#11 (338--752) & \#16 (602--1032) & \#5 (520--935) \\
DOLoRes@TNG\,3.6m   & La Palma, Spain   &         LR-B (300--843) & LR-R (447--1007) & \\
CAFOS@2.2m          & Calar Alto, Spain &           B200 (320--900) & R200 (620--1100) & \\
BFOSC@Cassini\,1.5m & Loiano, Italy     &             \#3 (330--642) &   \#5 (480--980) & \\
  \hline
  \end{tabular}
  \caption{Summary of the facilities used for the SPSS spectroscopy observations.}\label{tab:tels}
 \end{center}
\end{table}

\subsection{Spectra data reduction}

The absolute photometry of the SPSS was recently presented in [\citenum{altavilla20}], in this work we will focus on the spectroscopy.
Table~\ref{tab:tels} summarizes the instruments and spectral coverage used.
The data reduction is divided in two parts: a basic spectra reduction and an advance spectra reduction.

\begin{itemize}
    \item 
Basic Reduction.
First of all the data have been pre-reduced: correction for bias,  flat-field, bad pixel mask and illumination have been applied [\citenum{altavilla15}]. Then, the spectra have been extracted and wavelength calibrated. To avoid flux losses, we observed the spectra with wide slits (10--12"), but the seeing conditions influenced the resulting spectra. A spectrum was considered wide-slit if the ratio between the slit and the seeing during the observations was higher than 6, while it was considered narrow-slit if it was between 1.5 and 6, and it was only used only for wavelength calibration purposes if it was below 1.5. 
\item Advanced Reduction.
Some instrumental effects were present in the extracted spectra and additional reduction steps were applied to remove or mitigate them.
In particular, the red spectra used were affected by fringing. We used the procedure described in [\citenum{altavilla15}] to mitigate this effect. Moreover, the spectra (both blue and red) acquired with a narrow slit have been treated to remove the flux losses (Pancino et al. in preparation). As a last instrumental effect, we mention the second order contamination present in the spectra acquired with the grism 16 of EFOSC@NTT and in the redder part of the spectra acquired with grism LR-R of DOLoRes@TNG,
where light from the blue side of the second order contaminates the red side of the first order spectra. We corrected for this effect as described in [\citenum{altavilla15}].  
When these instrumental effects were corrected, we removed the telluric absorption bands using IRAF\footnote{\url{http://iraf.noao.edu}} ([\citenum{irafold}],[\citenum{iraf}]) and Molecfit ([\citenum{molecfit}],[\citenum{molecfit2}]) (Pancino et al. in preparation). 
\end{itemize}

\section{SPSS V2 Release}
\label{sec:v2}
\begin{figure}[ht!]
       \centering
       \includegraphics[width=.7\columnwidth,angle=0]{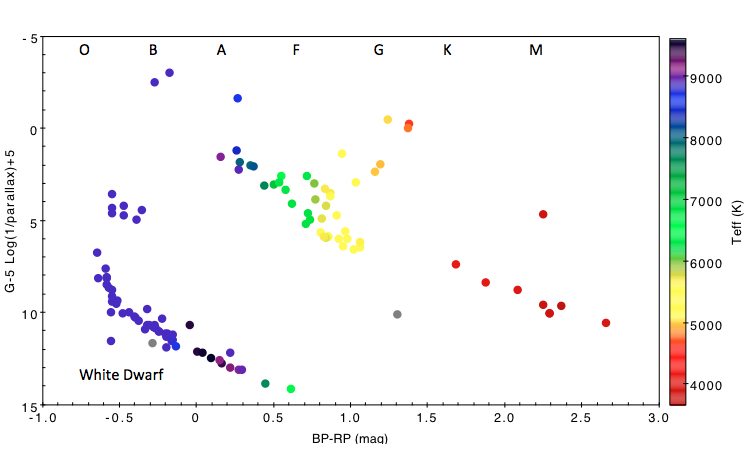}
       \caption{Distance-corrected CMD made with {\it Gaia} DR2 data of the 113 SPSS.}
        \label{fig:SpecTy}
\end{figure}

The current version (V2) of the grid is a preliminary release and will be used to  calibrate the {\it Gaia} (E)DR3 spectra and photometry. The final version, containing the entire SPSS sample ($\sim$ 200 stars), will be used to calibrate {\it Gaia} DR4.
The previous preliminary version (V1) used to calibrate the {\it Gaia} DR2 data contained a few stars discovered to be variable ([\citenum{marinoni16}]). These stars have been removed from the V2 while new stars of various spectral types, including two M stars, have been added, covering all spectral types, as shown in Figure \ref{fig:SpecTy}. The 
V2 release contains the flux tables, one per star, of 113 SPSS.

The choice of these stars is based on the spectral types (in order to include from O to M and many white dwarfs) and the quality of the spectra (observed in 26 photometric or grey nights, Pancino et al. in preparation).

\subsection{Calibration and flux tables preparation}

After the telluric contamination correction, the spectra have been calibrated in flux.
This procedure requires different steps:
\begin{enumerate} 
\item relative (nightly) flux calibration. All the spectra observed in the same night with the same instruments have been calibrated, using the extinction and response curves for that night and telescope; 
\item absolute flux calibration. All the wide spectra observed on photometric nights of the same SPSS have been used to build the median spectrum, on which we scaled all the lower quality spectra obtained in step 1, such as narrow-slit spectra or spectra observed on non-photometric nights; 
\item flux tables creation. All the spectra of the same SPSS calibrated in step 2 have been merged (after the rejection of the bad ones) using the median and the MAD (median absolute deviation) to obtain the final flux tables and their errors.   
\end{enumerate}

\subsection{Extension with templates}

All the flux tables were extended with template spectra to reach about 300~nm at the bluest wavelengths and  at least 1100~nm (possibly 1200~nm) at the reddest ones.
We selected theoretical models from public libraries provided by [\citenum{koester}], [\citenum{levenhagen}], [\citenum{coelho}], [\citenum{husser}] and the CALSPEC templates\footnote{\url{http://www.stsci.edu/hst/instrumentation/reference-data-for-calibration-and-tools/astronomical-catalogs/calspec}}. We chose the best-fitting model with a $\chi$-square minimization algorithm.

\section{Validation} 
\label{sec:val}

In order to validate the current release, a series of tests have been carried out, comparing the SPSS V2 and CALSPEC spectra, as well as absolute and synthetic photometry and the stellar parameters of the templates used for the extension and the literature.
CALSPEC is the library of flux standard stars built with HST data to calibrate the James Webb Telescope data [\citenum{bohlin14}]; [\citenum{bohlin19}]. The last CALSPEC release includes 90 stars, but only a few K and M-type stars.  

\begin{figure}[ht!]
       \centering
       \includegraphics[width=1.\columnwidth,angle=0]{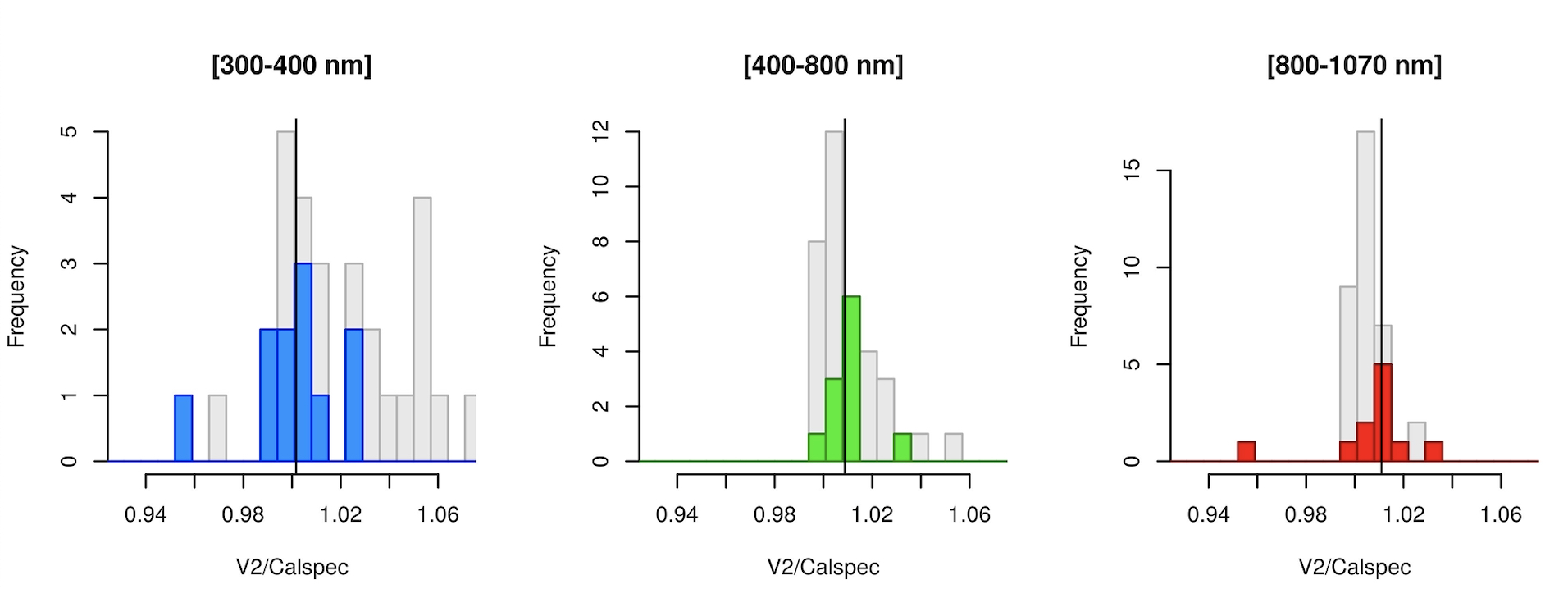}   
       \caption{Comparison between SPSS and CALSPEC spectra in different regions (blue: 300 - 400 nm; green: 400 - 800 nm; red: 800 - 1070 nm) for the V2 release. The internal CALSPEC spread is plotted in grey.}
        \label{fig:histo}
\end{figure}

\subsection{Spectra comparisons}

First of all, we compared the SPSS V2 spectra with the previous release. As expected, the overall agreement is extremely good proving that the two releases are on the same photometric system. In particular, we found that in the central and red regions of the spectrum (400-800~nm; 800-1070~nm, respectively) the V2 release is consistent with the previous one, while an improvement in the quality of the V2 spectra is visible in the blue (300-400~nm), given the increased number of spectra included in each final SPSS flux table.

We then compared the SPSS V2 spectra with CALSPEC.
We chose the latest CALSPEC version which contains observed and model spectra for 36 stars in common with the full SPSS set of 200 stars. The ratio between the model and the observed spectra can be used to represent the internal uncertainty of the CALSPEC grid and it is shown in Figure~\ref{fig:histo} in grey.
We have computed the SPSS V2/CALSPEC ratio for the 11 stars in common and the median and the MAD in three different regions of the spectrum (300-400 nm, 400-800 nm, 800-1070 nm). 

The results are shown in Figure~\ref{fig:histo} and in Table~\ref{tab:rel}. The SPSS V2 release is compatible with the CALSPEC internal spread and is $\simeq 1 \%$ brighter than the current CALSPEC release. This is in part due to the CALSPEC revision of the flux scale downwards by about 0.6 \% [\citenum{bohlin14}].  The SPSS V2 release is in fact calibrated on the previous CALSPEC scale and is thus brighter. 

\begin{table}
\vspace{0.5 cm}
    \centering
     \begin{tabular}{|c|c|c|c|}
     \hline 
     \textbf{Release} & \textbf{$300 - 400~nm$} & \textbf{$400 - 800~nm$}&\textbf{$800 - 1070~nm$} \\\hline
    V2/CALSPEC&$1.0017 \pm 0.0118$&$1.0088 \pm 0.0036$&$1.0109 \pm 0.0060$\\\hline
    \end{tabular}
    \caption{Median and MAD values of the spectra comparisons in three different ranges.}\label{tab:rel}
\end{table}

\subsection {Synthetic magnitudes comparisons}

\begin{figure}
\vspace{0.5cm}
       \centering
       \includegraphics[width=0.49\columnwidth,angle=0]{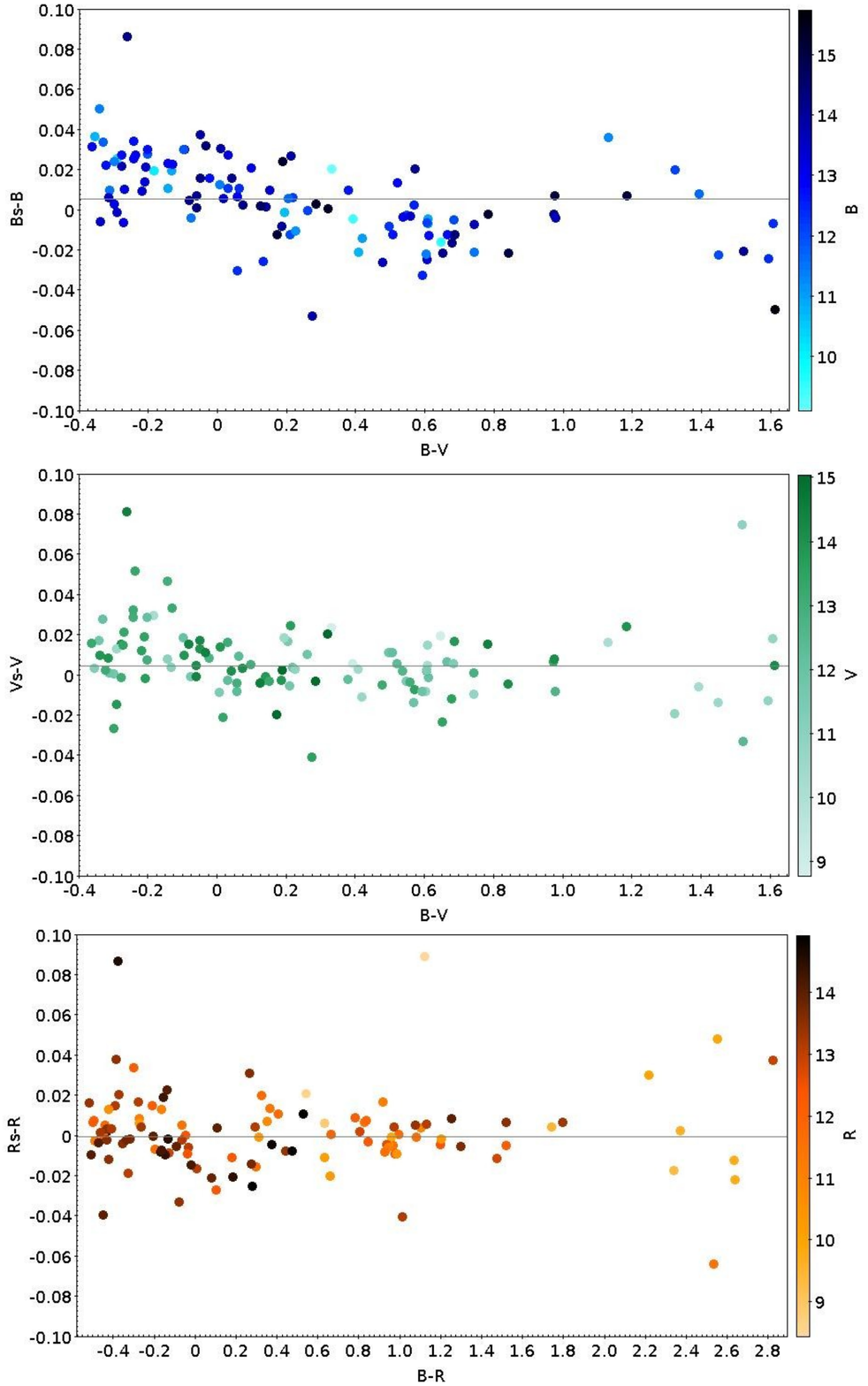}   
       \includegraphics[width=0.49\columnwidth,angle=0]{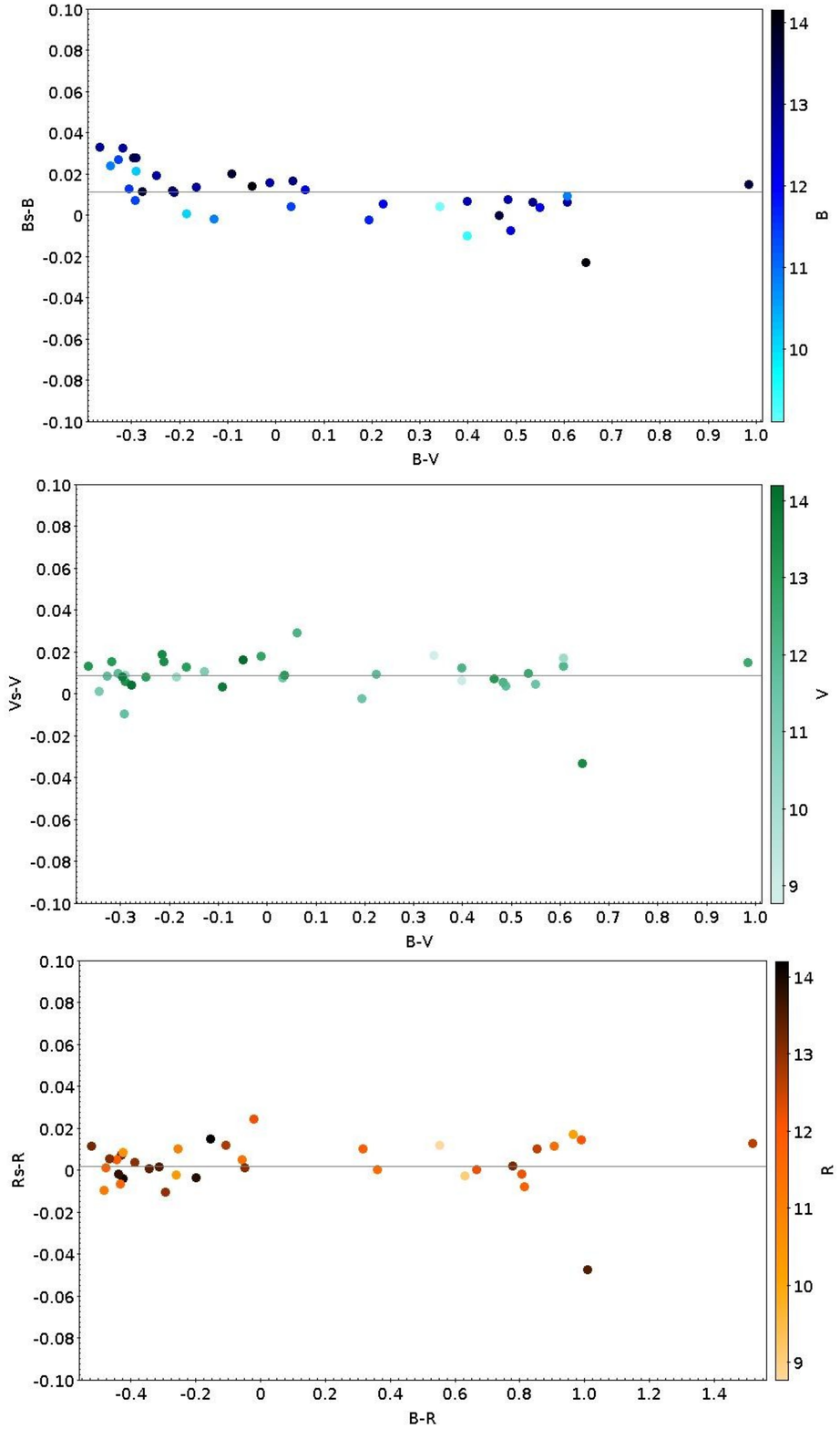}   
       \caption{Left: Comparison between synthetic photometry obtained using the SPSS V2 spectra and the SPSS absolute photometry for B (top), V (middle), R (bottom).
       Right: Comparison between synthetic photometry obtained using the SPSS V2 spectra and photometry by Landolt for B (top), V (middle), R (bottom).}
        \label{fig:magsin}
\end{figure}

We compared the synthetic Johnson-Kron-Cousins magnitudes derived from the SPSS V2 spectra, based on the [\citenum{bessell12}] passbands, with the magnitudes obtained in the SPSS absolute photometry campaign [\citenum{altavilla20}]. The left panels of Figure \ref{fig:magsin} show the cases of B (top), V (middle), R (bottom). The median values of $\Delta mags$ are in good agreement (see Table~\ref{tab:mag}), suggesting that the SPSS spectra are in the same photometric system as the SPSS absolute photometry. A slight trend of $\Delta B$ as a function of B-V can be noticed for blue stars in the top left panel. The trend is most likely caused by differences in the standard Landolt system, based on photoelectric photometry, and modern CCD measurements, as discussed estensively by [\citenum{clem13}],[\citenum{clem16}] and [\citenum{altavilla20}]. 
We tried different passbands from the literature and the effect varies in strength but never disappears. The chosen set of passbands minimized the slope in all bands. 

\newpage
\begin{table}
 \centering
 \begin{tabular}{|c|c|c|}
 \hline 
 \textbf{Band} & \textbf{Synthetic -- Absolute} & \textbf{Synthetic -- Landolt} \\\hline
 B& $0.0053 \pm 0.0145$&$0.0110 \pm 0.0072$\\\hline
 V&$0.0045 \pm 0.0096$&$0.0087 \pm 0.0044$\\\hline
 R&$-0.0007 \pm 0.0077$&$0.0016 \pm 0.0067$\\\hline
  \end{tabular}
 \caption{Median and MAD values of the photometries comparisons.}
 \label{tab:mag}
\end{table}

We also compared the synthetic magnitudes derived from the SPSS V2 spectra with the literature values derived by [\citenum{landolt92a}] and a collection of papers by Landolt and collaborators, assembled as described by [\citenum{altavilla20}], for the 37 stars in common.
The results are shown in the right panels of Figure \ref{fig:magsin}. The median values of $\Delta mags$ are in very good agreement (see Table~\ref{tab:mag}) for the R band, but our SPSS synthetic magnitudes are slightly fainter ($\simeq 1\%$) than the reference Landolt ones in B and V.

\subsection{Parameters comparisons}

\begin{figure}[ht!]
       \centering
       \includegraphics[width=1\columnwidth,angle=0]{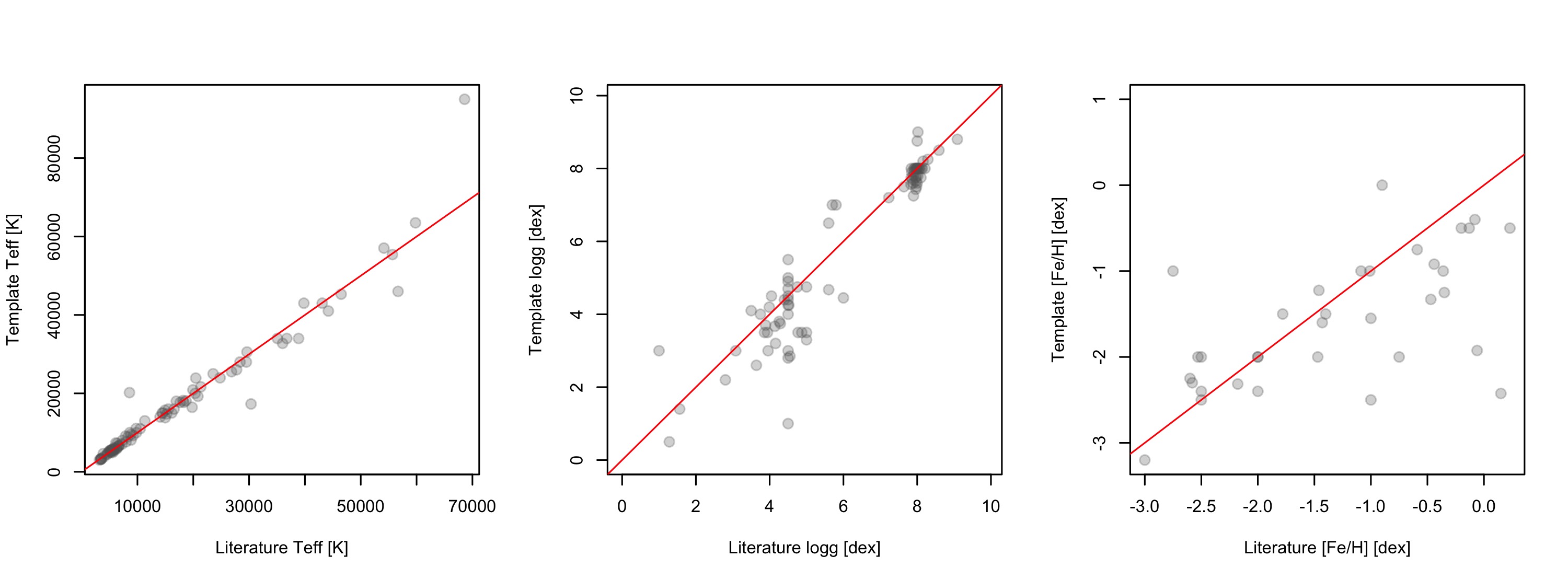}   
       \caption{Stellar parameters comparison. Temperatures (left) and gravities (middle) of the models used to extend the SPSS spectra are in good agreement with the literature ones. This is less  true for the metallicities (right).}
        \label{fig:params}
\end{figure}

In order to check the template choice for the SPSS V2 spectra extension, we have also compared the stellar parameters of the best-fitting models used for this scope with the values available for each SPSS in the literature.
Figure \ref{fig:params} shows the comparison between temperature, gravity, and metallicity. The median differences (and MAD) are:
$\Delta T_{eff} = -60$ ($\pm 697$) K, $\Delta\log g = -0.14$ ($\pm 0.33$) dex, $\Delta [Fe/H] = -0.17$ 
($\pm 0.56$) dex .
The temperatures and gravities are in good agreement, while this is less true for the metallicities. This is expected considering that the SPSS spectra have low resolution  (tipically $R = \lambda / \Delta \lambda \simeq 500-1000$) and many of our SPSS have high surface temperaturs, and thus they are not very sensitive to [Fe/H]. Moreover, most of the spectra used to derive the metallicity in the literature are low resolution spectra as well, thus the literature [Fe/H] values suffer from similar uncertainties. Anyway, the overall agreement of the parameters is satisfactory for the purpose of the present work.

\section{Summary and Conclusions}
\label{sec:conc}
We presented the preliminary version (V2) of the SPSS grid used to calibrate the {\it Gaia} (E)DR3 photometric and spectroscopic data. It consists on flux tables of 113 stars. The sample includes different spectral types (from O to M and many white dwarfs) covering the entire {\it Gaia} spectral range ($\sim$ 300--1100/1200~nm) with an internal precision and an external accuracy of $\simeq$~1\%.  We validated the grid performing different tests on the spectra, their synthetic photometry, and the stellar parameters. For the spectra validation we compared SPSS V2 and the CALSPEC spectra founding the SPSS brighter of $\sim$1~\% than CALSPEC. 
Regarding the photometry, we first compared the synthetic Johnson-Kron-Cousins magnitudes derived from the SPSS V2 spectra with the SPSS absolut photometry, finding that the data are in the same phototmetric system. Then we compared the synthetic photometry with the Landolt photometry. We found that the SPSS are $\sim$1~\% fainter than Landolt. This means that the SPSS are in between the two best standard datasets available in the literature. As final test we compared the stellar parameters of the model used to extend the spectra with the available literature's values for the SPSS, finding that temperatures and gravties are in very good agreement.

The quality of the SPSS flux tables already meets the {\it Gaia} formal requirements, but it is expected to further improve in future releases, that will be used for the calibration of the final {\it Gaia} release (DR4), when a larger number of reduced spectra (about 6500 rather than 1500) and a larger number of SPSS ($\sim$200) will be used.

\acknowledgments 
This work has used data from the European Space Agency (ESA) mission \textit{Gaia}, processed by the \textit{Gaia} Data Processing and Analysis  Consortium (DPAC, \url{https://www.cosmos.esa.int/web/gaia/dpac/consortium}). Funding for the DPAC  has  been  provided  by  national  institutions, in particular the institutions participating in the \textit{Gaia} Multilateral Agreement. The following software have been used for doing the figures and the statistical analysis: the R programming language and Rstudio (https://www.rstudio.com/); Topcat catalogue plotting tool ([\citenum{taylor}]).


\bibliography{report.bib} 
\bibliographystyle{spiebib.bst} 

\end{document}